\documentclass[12pt]{article}

\usepackage{lineno}
\RequirePackage{amsthm,amsmath,amsfonts,amssymb}
\usepackage{natbib}
\usepackage{amsmath}
\usepackage{graphicx,psfrag,epsf}
\usepackage{enumerate}
\usepackage{xcolor}
\usepackage{natbib}
\usepackage{multirow}
\usepackage{multicol}
\usepackage{booktabs}
\usepackage{url}
\usepackage{verbatim}
\usepackage{tablefootnote}
\usepackage{bbm}
\usepackage{mathrsfs}
\usepackage{natbib,color}
\usepackage{amsfonts}
\usepackage{amsmath,amssymb,amsthm}
\usepackage{physics}
\usepackage[colorlinks,citecolor=blue,linkcolor=blue,urlcolor=blue,pdfborder={0 0 0}]{hyperref}
\usepackage{xcolor}
\usepackage{makecell}
\usepackage{booktabs} 
\usepackage{enumitem}

\newtheorem{proposition}{Proposition}

\newtheorem{theorem}{Theorem}
\newtheorem{remark}{Remark}

\newcommand{\blind}{1}

\def\P{\mathbf{P}}
\def\W{\mathbf{W}}

\def\S{\mathbf{S}}
\def\A{\mathbf{A}}
\def\B{\mathbf{B}}

\def\W{\mathbf{W}}

\def\R{\mathbb{R}}

\def \E {\mathrm{E}}

\def \tr {\operatorname{tr}}

\begin{document}

\def\spacingset#1{\renewcommand{\baselinestretch}%
{#1}\small\normalsize} \spacingset{1}

\if1\blind
{
  \title{\bf WISE: A Weighted Similarity Aggregation Test for Serial Independence}
  
  \author{
    Qihua Zhu$^{1}$\thanks{Zhu and Liu contributed equally to this work.},
    Mingshuo Liu$^{2}$\footnotemark[1],
    Yuefeng Han$^{3}$,
    Doudou Zhou$^{1}$\thanks{Corresponding author. Email: \texttt{ddzhou@nus.edu.sg}.} \\[1ex]
    {\footnotesize $^{1}$Department of Statistics and Data Science, National University of Singapore} \\
    {\footnotesize $^{2}$Department of Statistics, University of California, Davis} \\
    {\footnotesize $^{3}$Department of Applied and Computational Mathematics and Statistics, University of Notre Dame}
  }

  \date{}
  
  \begingroup
    \renewcommand\thefootnote{\fnsymbol{footnote}}
    \maketitle
  \endgroup
  \setcounter{footnote}{0}
  \renewcommand\thefootnote{\arabic{footnote}}

} \fi

\if0\blind
{
  \bigskip
  \bigskip
  \bigskip
  \begin{center}
    {\LARGE\bf WISE: A Serial Independence Test Based on Weighted Information Similarity Aggregation}
\end{center}
  \medskip
} \fi


\bigskip
\begin{abstract}
We propose a nonparametric test for serial independence that aggregates pairwise similarities of observations with lag-dependent weights. The resulting statistic is powerful to general forms of temporal dependence, including nonlinear and uncorrelated alternatives, and applies to ultra-high-dimensional and non-Euclidean data. We derive asymptotic normality under both permutation and population nulls, and establish consistency in classical large-sample and high-dimension-low-sample-size (HDLSS) regimes. The test therefore provides the first theoretical power guarantees for serial independence in the HDLSS setting. Simulations demonstrate accurate size and strong power against a wide range of alternatives, showing significant power improvement over existing methods under various high-dimensional time series models. An application to spatio-temporal data illustrates the method's utility for non-Euclidean observations.
\end{abstract}

\noindent%

\newpage
\spacingset{1.8} 

\section{Introduction}
\label{sec:intro}

Let $\{X_t\}_{t=1}^n$ be a sequence of $\mathcal{X}$-valued random variables sharing a common marginal distribution $G$, where $\mathcal{X}$ is a measurable space sufficiently general to accommodate complex data types. The problem of serial independence concerns testing
\begin{equation*}
\mathsf{H}_0:\ \{X_t\}_{t=1}^n \text{ are i.i.d.} \quad\text{vs.}\quad
\mathsf{H}_a:\ \{X_t\}_{t=1}^n \text{ are not independent.}
\end{equation*}
This classical problem has been extensively studied due to its fundamental role in time series analysis. On the theoretical side, many diagnostic testing problems for time series models, such as tests for linearity or for the random walk hypothesis, can be reformulated as tests of serial independence \citep{robinson1991consistent, Hjellvik1996NonparametricSF, hong1998testing}. On the applied side, serial independence tests have been used to assess cointegration relationships \citep{Chigira2006ATO}, to examine the temporal dependence of stock returns \citep{ZhouTestingHW}, and to investigate the predictability of financial markets \citep{Ashley2024TheAA}.

Most existing work on serial independence focuses on white-noise or autocorrelation tests. In both univariate and multivariate contexts,  classical approaches can be divided into \textit{time-domain methods}, based on the autocorrelation function, and \textit{frequency-domain methods}, based on the spectral density. Time-domain examples include the Box-Pierce and Ljung-Box tests \citep{box1970distribution, ljung1978measure}, which laid the foundation for subsequent developments \citep{davies1979some, hosking1980multivariate, lobato2001testing, escanciano2009automatic, wei2024testing}. Frequency-domain methods, such as those of \cite{hong1996consistent}, \cite{deo2000spectral}, and \cite{shao2011testing}, assess deviations of the sample spectral density function from flatness and are widely used for testing white noise under weak assumptions on dependence.

In high-dimensional statistics, recent efforts have developed procedures for testing white noise hypotheses \citep{chang2017testing, li2019testing, Tsay2020TestingSC, Chen2022RankBT}. Max-type tests \citep{chang2017testing} are generally effective when correlations are sparse, while sum-type test \citep{li2019testing} suits dense correlations. Distribution-free approaches based on rank correlations \citep{Tsay2020TestingSC, Chen2022RankBT} offer robustness to heavy tails, and self-normalization methods can also be adapted \citep{Wang2020HypothesisTF}. These approaches are powerful for detecting linear dependence but often lack sensitivity to uncorrelated but dependent alternatives, such as Autoregressive Conditional Heteroskedasticity (ARCH), Generalized Autoregressive Conditional Heteroskedasticity (GARCH), and some Nonlinear Moving Average (NMA) models \citep{tong1990non}.

To address more general forms of serial dependence, several nonparametric approaches have been proposed. For univariate data, these include tests based on the correlation integral \citep{brock1987test}, empirical distribution function \citep{skaug1993nonparametric, hong1998testing}, generalized spectral function \citep{ hong1999hypothesis, hong2000generalized}, and distance covariance \citep{zhou2012measuring, fokianos2017consistent}. Extensions to multivariate settings have also been studied \citep{baek1992nonparametric, fokianos2018testing}. Most recently, \cite{jiang2024testing} introduced a distance-based Cramér-von Mises-type test applicable to both Euclidean and non-Euclidean data.

\begin{figure}[t]
    \centering \includegraphics[width=0.85\linewidth]{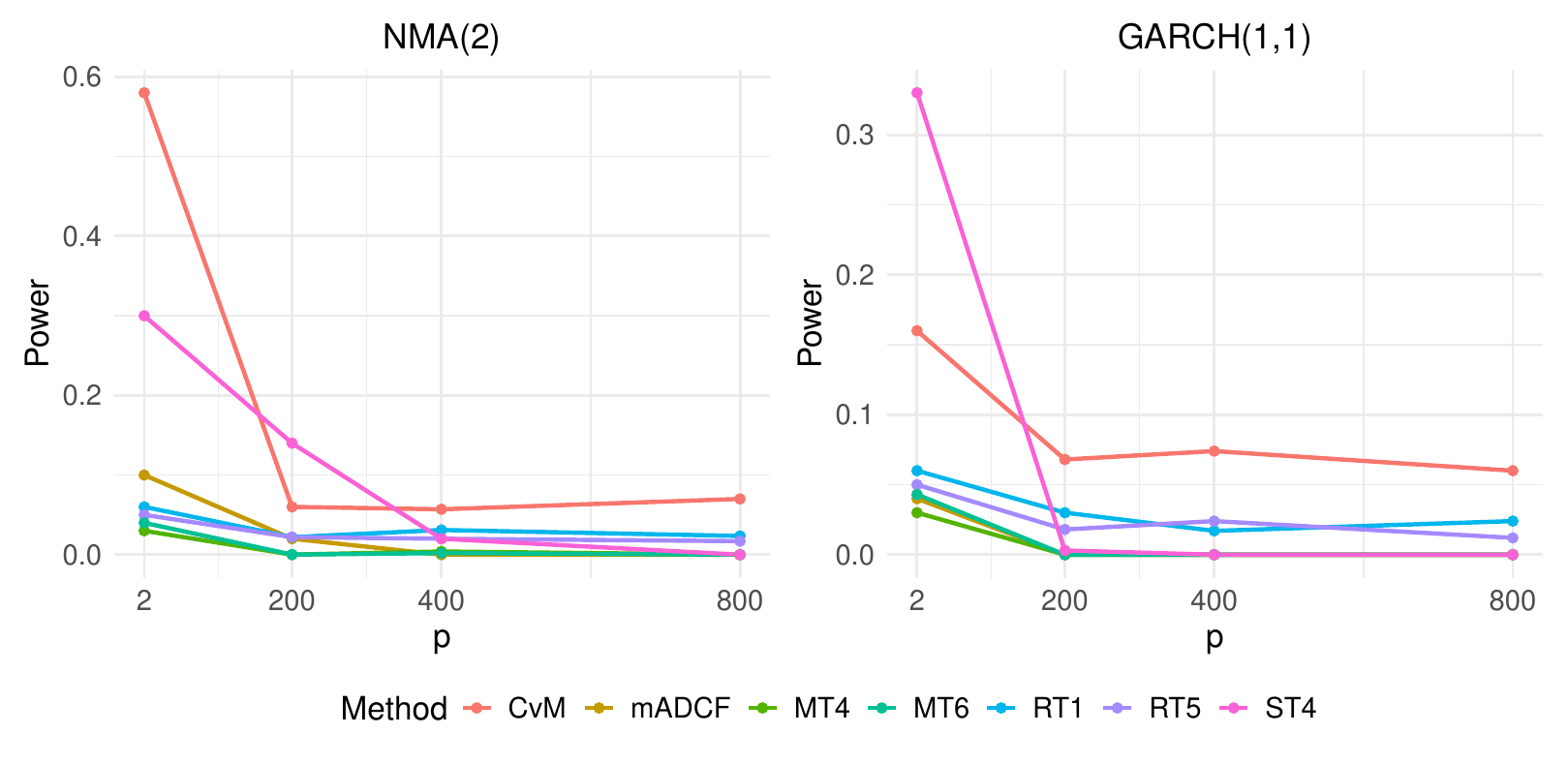}
    \caption{Empirical power of five recent tests under two uncorrelated but dependent alternatives. Left: NMA(2), $X_t = \epsilon_t \circ \epsilon_{t-1} \circ \epsilon_{t-2}$, where $\circ$ denotes the Hadamard product. Right: multivariate GARCH(1,1),  $X_{t} = h_{t}\circ \epsilon_{t}$, with $h_t^{\circ 2}= b + \A X_{t-1}^{\circ 2} + \B h_{t-1}^{\circ 2}$, $x^{\circ 2}$ representing $ x \circ x $, $b = (0.002,\dots,0.002)\in \R^p$, and $\A$ and $\B$ being the diagonal matrices.  For $p=2$, diagonal entries $A_{ii},B_{ii}$ are independently drawn from $\mathrm{U}(0.5,0.6)$ and $\mathrm{U}(0.3,0.4)$, respectively; for $p>2$, they are drawn from $\mathrm{U}(0,0.15)$ and $\mathrm{U}(0,0.4)$.   In both settings $\epsilon_t \overset{i.i.d.}{\sim} N(0,\mathbf{I}_p)$.  }
    \label{fig:dataexample}
\end{figure}

Despite these advances, current methods remain limited in their ability to detect uncorrelated serial dependence in high-dimensional or non-Euclidean data. To illustrate, Figure~\ref{fig:dataexample} compares five recent tests under two uncorrelated alternatives: a NMA model and a multivariate GARCH process. The competitors include max-type tests (MT4, MT6, \citealp{chang2017testing}) designed for sparse strong correlations; the multivariate auto-distance correlation test (mADCF, \citealp{fokianos2018testing}); a sum-type test (ST4, \citealp{li2019testing}) suited for dense correlations; rank-based tests (RT1, RT5, \citealp{Tsay2020TestingSC}) robust to heavy tails; and a Cram\'er-von Mises-type statistic based on auto-distance covariance (CvM, \citealp{jiang2024testing}). We consider $n=100$, dimensions $p\in\{2,200,400,800\}$, and nominal level $0.05$, with results averaged over $150$ replications. While CvM and ST4 achieve moderate power in low dimensions, all methods experience severe power loss once $p>n$, highlighting the need for new approaches capable of capturing complex dependence in high-dimensional time series.

To address these challenges, we propose a new framework for testing serial independence, termed \textbf{W}eighted \textbf{I}nformation \textbf{S}imilarity aggr\textbf{E}gation (WISE). The method constructs a similarity matrix whose entries quantify pairwise similarities between observations, and then aggregates these with lag-dependent weights. Under $\mathsf{H}_0$, the similarity field is approximately homogeneous, while departures from independence generate structured deviations that are amplified by the weighting scheme.

WISE has several attractive features. First, unlike most permutation-based procedures that require intensive resampling, WISE admits analytic null moments and a normal approximation under the permutation distribution, greatly improving scalability. This approximation holds under mild conditions on the similarity structure, without moment assumptions or restrictions on the growth of $n$ and $p$. We also establish asymptotic normality under the population null, requiring only mild moment conditions. Second, the test is theoretically supported: it is consistent under both the conventional large-sample regime and the HDLSS regime with $n$ fixed and $p\to\infty$, providing, to our knowledge, the first power analysis of a serial independence test in the HDLSS setting. Simulations further demonstrate strong performance in ultra-high dimensions, particularly against uncorrelated but dependent alternatives. Third, WISE naturally extends to non-Euclidean data whenever an appropriate similarity measure is available, whereas among existing methods only the procedure of \citet{jiang2024testing} applies directly in such settings.

The remainder of the paper is organized as follows. Section~\ref{sec:Method} introduces the methodology, Section~\ref{sec:theroy} establishes theoretical properties, and Section~\ref{sec:Performance} reports simulation results. Section~\ref{sec:RealData} illustrates the method with spatio-temporal data, and Section~\ref{sec:conc} concludes. Proofs and additional experiments are given in the Supplementary Material.

\section{Method}\label{sec:Method}
We first describe the construction of the similarity matrix in Section \ref{sec:sim-matrix}, then introduce the test statistic in Section \ref{sec:test stat}. Guidelines for choosing lag-dependent weighting schemes, aimed at improving power in practice, are given in Section \ref{sec:weight schemes}.

\subsection{Similarity matrix}\label{sec:sim-matrix}

The basic building block of our procedure is a pairwise similarity function $S:\mathcal{X}\times\mathcal{X}\to\mathbb{R}$ with $S(X_i,X_j)=S(X_j,X_i)$. If $S$ is not symmetric, we replace it by $\{S(X_i,X_j)+S(X_j,X_i)\}/2$. The quantity $S(X_i,X_j)$ measures how similar two observations are and thus serves as a local dependence proxy for $\{X_t\}_{t=1}^n$. Intuitively, if $X_i$ and $X_j$ are dependent, their similarity tends to be systematically higher (or lower) than that of independent pairs. Common choices include distance-based measures (e.g., negative Euclidean or $\ell_1$ distance), graph-based measures (e.g., $k$-nearest-neighbor affinities), and kernel-based measures (e.g., Gaussian kernels). Additional discussion appears in Supplementary~A.

We aggregate these local comparisons in the $n\times n$ similarity matrix $\S=[S_{ij}]_{i,j=1}^n=[S(X_i,X_j)]_{i,j=1}^n$, which acts as a matrix-valued empirical dependence summary. Unlike scalar measures (e.g., autocorrelations or distance covariances) that encode dependence through magnitude, $\S$ also reveals dependence through structure. Under $\mathsf{H}_0$, the off-diagonal entries $\{S(X_i, X_j):i\neq j\}$ are identically distributed, so $\S$ should be approximately homogeneous (aside from its diagonal). Under dependence, $\S$ tends to develop systematic patterns, such as bands, blocks, or clusters that encode temporal structure.

\begin{figure}[!t]
    \centering
    \includegraphics[width=1\linewidth]{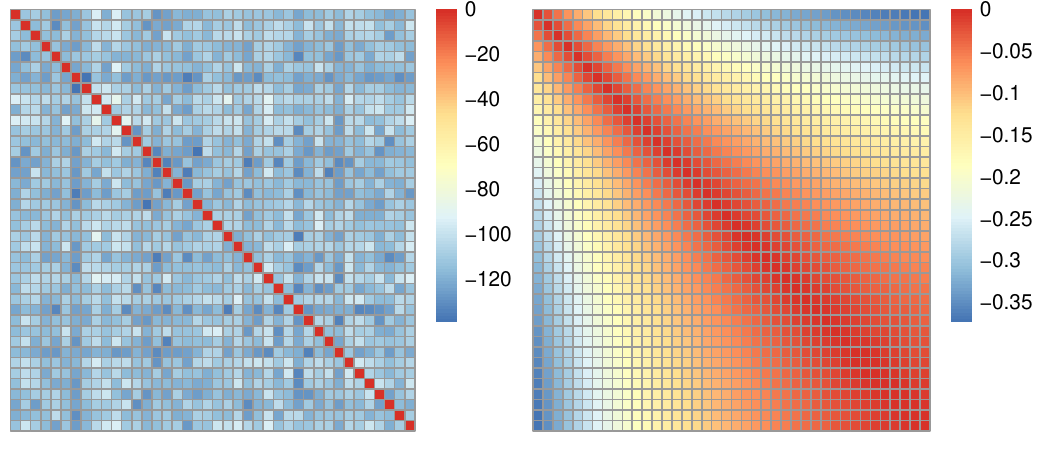}
    \caption{Heatmaps of $\S_{\mathrm{I}}$ (left) and $\S_{\mathrm{II}}$ (right) with $n = 40$, $p = 100$ and $S(X_i, X_j) = - \|X_i - X_j\|_{1} =  -\sum_{k=1}^p |X_{ik} - X_{jk}|$.}
    \label{fig:heatmaps for similarity matrix}
\end{figure}

To illustrate, consider (I) $X_t \overset{\text{i.i.d.}}{\sim} N(0,\mathbf{I}_p)$ and (II) a VAR(3) sequence $X_t=0.4X_{t-1}+0.3X_{t-2}+0.2X_{t-3}+\epsilon_t$ with $\epsilon_t\overset{\text{i.i.d.}}{\sim}N(0,\mathbf{I}_p)$. Figure~\ref{fig:heatmaps for similarity matrix} displays heatmaps of $\S$ in both settings with $n=40$, $p=100$, and $S(X_i,X_j)=-\|X_i-X_j\|_1$. Under independence, $\S$ is nearly pattern-free away from the diagonal, while the VAR process produces a pronounced banded structure with large values near the diagonal, reflecting stronger similarity at small lags.

\subsection{Test statistic}\label{sec:test stat}
To quantify these structural deviations, we form a weighted aggregation of the similarity field:
\begin{equation} 
\label{eqn:double indexed} 
Z = \sum_{i=1}^n \sum_{j=1}^n w(|i-j|) S(X_i, X_j), 
\end{equation} 
where $w: \mathbb{R} \rightarrow \mathbb{R}$ is a user-specified weight function with $w(0) = 0$. We reject $\mathsf{H}_0$ when $|Z|$ is large relative to its permutation null distribution, obtained by recomputing \eqref{eqn:double indexed} on $\{X_{\pi(t)}\}_{t=1}^n$ for uniformly sampled permutations $\pi$ of $[n] = \{1,\ldots,n\}$. The statistic exploits the fact that under $\mathsf{H}_0$ permutations merely reshuffle a homogeneous field, whereas under dependence \emph{lag-structured} deviations are amplified by suitable choices of $w$. In practice, our method allows users to choose various weighting schemes according to the alternative of their interest, which can serve as a way to enhance the power performance of the test. Section~\ref{sec:weight schemes} provides guidelines for choosing weight functions tailored to specific alternatives, thereby improving power.

Exact permutation critical values are infeasible for large $n$. We therefore standardize $Z$ as
\begin{equation}
    Z_G = \frac{Z - \mathrm{E}(Z)}{\sqrt{ \mathrm{var}(Z) }},
    \label{eqn:test_stat} 
\end{equation}
where $\mathrm{E}$ and $\mathrm{var}$ are taken under the permutation distribution. Closed-form expressions for $\mathrm{E}(Z)$ and $\mathrm{var}(Z)$ are given in Theorem~\ref{thm:expect,cov}. Under mild conditions, $Z_G$ converges in distribution to $N(0,1)$ under the permutation null (Theorem~\ref{theory:normaltity}), so we reject $\mathsf{H}_0$ when $|Z_G|>z_{1-\alpha/2}$, where $z_{1-\alpha/2}$ is the $(1-\alpha/2)$-quantile of the standard normal distribution.

\begin{remark}[Non-Euclidean data]\label{re:sim4non-euc}
The framework applies directly to non-Euclidean data once a similarity is specified.
Examples include: 
    \begin{itemize}
        \item[(a)] Matrix-valued data  $\Omega = \mathbb{R}^{p\times p}$ with  $S(X_i,X_j) = -\|X_i-X_j\|_\mathrm{F}$, where $\|\cdot\|_\mathrm{F}$ denotes the Frobenius norm. This setting arises when $X_i$ represents, e.g., an adjacency matrix of a network or a pixel matrix of an image.
        
        \item[(b)] Functional data $\Omega=\{f:[0,1]\to\mathbb{R}:\int_0^1 f^2(\tau)\,d\tau<\infty\}$ with 
        $$
        S(X_i,X_j)=-\Bigl(\int_0^1 (X_i(\tau)-X_j(\tau))^2\,d\tau\Bigr)^{1/2}. 
        $$
        
        \item[(c)] Distributional data: $\Omega$  the set of cumulative distribution functions on $[0,1]$ with
        $$
        S(X_i,X_j)=-\int_0^1 |X_i^{-1}(\tau)-X_j^{-1}(\tau)|\,d\tau,
        $$
        where $X_i^{-1}$ and $X_j^{-1}$ denote quantile functions. 
    \end{itemize}
  Section~\ref{sec:RealData} illustrates performance on spatio-temporal data.
\end{remark}

\subsection{Choice of weight functions}\label{sec:weight schemes}

While $Z$ is centered near $\mathrm{E}(Z)$ under $\mathsf{H}_0$ regardless of $w$, its departure under alternatives depends critically on the weighting. The following result motivates how to select $w$.
\begin{proposition}\label{prop: weight selection}
Let 
$S^{(1)}\leq \dots \leq S^{(n^2)}$ and 
$w^{(1)}\leq \dots \leq w^{(n^2)}$ be 
the ordered entries of the similarity matrix $\S=[S_{ij}]_{i,j=1}^n$ and the weight matrix $\W=[w_{ij}]_{i,j=1}^n=[w(|i-j|)]_{i,j=1}^n$. For any permutation $\pi$ of $[n]$, 
$$
\sum_{k=1}^{n^2} w^{(k)} S^{(n^2-k+1)}
\;\leq\; 
Z_\pi
\;\leq\;
\sum_{k=1}^{n^2} w^{(k)} S^{(k)},
$$
where $Z_\pi = \sum_{i=1}^n \sum_{j=1}^n w(|i-j|) S_{\pi(i)\pi(j)}$ is the value of $Z$ computed on a permutation $\pi$ of the indices.
\end{proposition}
Proposition~\ref{prop: weight selection} shows that pairing larger similarities with larger (or smaller) weights yields a more extreme $Z_\pi$ relative to its permutation counterparts, increasing power. This suggests the general principle:
\emph{assign larger weights to entries of $\S$ that are expected to be systematically larger or smaller under the alternative of interest.}
Different alternatives induce different value patterns in $\S$, and $w$ should reflect those patterns. We outline three common cases.

\vspace{10pt}
\noindent \textbf{Proximity-type.} A common and practically important structure is proximity-type dependence, where the strength of dependence between two observations typically decreases as their time gap increases. This pattern arises in many classical models, including first-order autoregressive  (AR(1)) process, the first-order moving average (MA(1)) process, the exponential smoothing process, and the first-order autoregressive conditional heteroskedasticity (ARCH(1)) process. In such contexts, it is natural to use a monotone weight function with respect to $|i-j|$. Representative choices include: algebraic decay 
$w(t) = (1+t)^{-\beta} - 1$, with $\beta>1$; geometric decay $w(t) = \rho^{t} - 1$ with $\rho \in (0,1)$; and exponential decay $w(t) = \exp(-(t/\lambda)^2) - 1$, with $\lambda>0$. We subtract $1$ so that $w(0)=0$ and the weights are negative away from the diagonal, consistent with our centering scheme. These families enable practitioners to tune the rate of decay to the expected dependence structure of the series. 

\vspace{10pt}
\noindent \textbf{Periodicity-type.} Many real-world series, such as temperature records, electricity demand, and air passenger volumes, exhibit  seasonal or periodic patterns. To target such alternatives, $w(|i-j|)$ should emphasize pairs of observations whose time gaps are multiples of a fundamental period $l$.  A natural strategy is to employ trigonometric functions, for example, $w(|i-j|) = \cos(2\pi |i-j|/l) - 1$ or $w(|i-j|) = \left|\cos\left(\pi |i-j|/l\right)\right| - 1$, where $l$ may be chosen using domain knowledge. If multiple periodic components are present, a Fourier-type weighting function can be used: $\sum_{k=1}^q\alpha_k\text{cos}(2\pi|i-j|/l_k)-1$, where $\alpha_k\in(0,1)$, $\sum_{k=1}^q\alpha_k = 1$ and $l_k$ denotes the period of the $k$th component. 

\vspace{10pt}
\noindent \textbf{Mixed-type.} Some processes combine proximity and periodic features. A typical example is the Seasonal Autoregressive Integrated Moving Average (SARIMA) process  (see Section 9.9 of \cite{hyndman2018forecasting}), which involves both proximity-type dependence and periodicity-type dependence. In such cases, $w(|i-j|)$ can be defined as a weighted combination of the individual weight functions.  For example, $w(|i-j|) = \alpha[(1/ |i-j|^\beta )- 1)]+(1-\alpha)[\text{cos}(2\pi|i-j|/l)-1]$ with $\alpha \in (0,1)$, where $\alpha$ controls the balance between proximity and periodicity.  

\vspace{10pt}
While many other forms of serial dependence arise in practice, it is impractical to enumerate them all. Our goal is to provide a general framework together with guidance for choosing the weighting function $w(\cdot)$ so as to enhance power. For concreteness, the subsequent theory and experiments focus on the simple proximity-oriented choice
\begin{equation}
\label{eqn:w(|i-j|)}
    w(|i-j|) = \frac{1}{1+(i-j)^2}-1,
\end{equation}
selected for its simplicity and the prevalence of proximity-type dependence in applications. Although tailored to proximity-type alternatives, Section~\ref{sec:Performance} shows that this weighting scheme also performs well across other forms of dependence. Accordingly, we use \eqref{eqn:w(|i-j|)} as the \emph{default} weight when domain-specific guidance is unavailable and throughout the remainder of the paper.

\section{Theoretical Properties}\label{sec:theroy}
This section establishes theoretical guarantees for the proposed test. In Section \ref{sec:Asymptotic normality}, we show that under the permutation null, the statistic in \eqref{eqn:test_stat} converges to a standard normal distribution as $n \to \infty$, validating normal critical values in large samples. In Section \ref{sec:consistency}, we analyze the power of WISE in both large-sample and HDLSS regimes. Throughout, for nonnegative sequences $a_n,b_n$, write $a_n\lesssim b_n$ if there is a constant $C>0$ (independent of $n$) such that $a_n\le C b_n$ for all  $n>0$, and $a_n \asymp b_n$ if $a_n\lesssim b_n$ and $b_n\lesssim a_n$.

\subsection{Asymptotic normality}\label{sec:Asymptotic normality}

We collect notation used throughout. Let 
\begin{align*}
& w_1 = \sum_{i=1}^{n} \sum_{j \neq i}^n  w(|i-j|), \quad w_{i \cdot} = \sum_{j \neq i}^n w(|i-j|), \text{ for } i \in \{1,\ldots,n\},   \\
&  w_2 = \sum_{i=1}^{n} \sum_{j \neq i}^n  w^2(|i-j|), \quad  w_3= \sum_{i=1}^{n} w_{i \cdot} w_{i \cdot}.
\end{align*}
We define $S_1,S_{i\cdot},S_2,S_3$ analogously by replacing $w(|i-j|)$ with $S(X_i,X_j)$.
\begin{theorem}
\label{thm:expect,cov}
  Under the permutation null distribution, we have,
$$
\E( Z ) = \frac{w_1 S_1}{n(n-1)},
$$
and
\begin{equation}
\begin{aligned}
\mathrm{var}( Z )= & 
\frac{4(n+1) (w_3-\frac{w_1^2}{n}) (S_3-\frac{S_1^2}{n}) }{n(n-1)(n-2)(n-3)}  + \frac{2(w_2-\frac{w_1^2}{n(n-1)}) (S_2-\frac{S_1^2}{n(n-1)})}{n(n-3)} \\
&  - \frac{4(w_2-\frac{w_1^2}{n(n-1)})(S_3-\frac{S_1^2}{n})}{n(n-2)(n-3)}   - \frac{4(w_3-\frac{w_1^2}{n})(S_2-\frac{ S_1^2}{n(n-1)})}{n(n-2)(n-3)}.  
\end{aligned}
\label{eqn: cov expression}
\end{equation}
Here $ \mathrm{E} $ and $ \mathrm{var} $ denote the expectation and variance under the permutation null distribution, calculated based on the permuted sequence $\{X_{\pi(t)}\}_{t=1}^n$, with $\pi$ uniformly sampled from all permutations of $[n]$.
\end{theorem}
Theorem \ref{thm:expect,cov} provides the exact expectation and variance under the permutation null. For asymptotics, it is convenient to work with centered quantities: 
$$
\tilde w(|i-j|):=w(|i-j|)-\frac{w_1\,\mathbf 1\{i\ne j\}}{n(n-1)},\qquad
\tilde S(X_i,X_j):=S(X_i,X_j)-\frac{S_1\,\mathbf 1\{i\ne j\}}{n(n-1)}.
$$
Since the standardized statistic in \eqref{eqn:test_stat} is invariant to this centering, we therefore define
\begin{equation}
\tilde Z=\sum_{i=1}^n\sum_{j=1}^n \tilde w(|i-j|)\,\tilde S(X_i,X_j),  
\label{eqn:centeredZ}
\end{equation}
and summarize $\widetilde\W=[\tilde w(|i-j|)]_{i,j=1}^n$ and $\widetilde\S=[\tilde S(X_i,X_j)]_{i,j=1}^n$ by
\begin{equation*}
\begin{aligned}
& \tilde w_{0+}=\max_{1\leq i,j \leq n} \big|\tilde w(|i-j|)\big|, ~ \tilde w_{i \star} = \sum_{j=1}^n \big|\tilde w(|i-j|)\big|, ~ i \in \{1,\ldots,n\}, \\
&  \tilde w_{1+} = \max_{1 \leq i \leq n} \tilde w_{i \star}, ~
 \tilde w_{2+}= \sum_{i=1}^{n} \sum_{j=1}^n  \tilde w^2(|i-j|), ~  \tilde w_{3+}= \sum_{i=1}^{n}  \tilde w_{i \star} \tilde w_{i \star},
\end{aligned}    
\end{equation*}
and analogously $\tilde S_{0+},\tilde S_{i\star},\tilde S_{1+},\tilde S_{2+},\tilde S_{3+}$.

\begin{theorem}
\label{theory:normaltity}
Assume that $\tilde \S$ satisfies
\begin{align}
\label{cond2,order}
\frac{n \tilde S_{0+}^2}{\tilde S_{2+}} \overset{\P}{\rightarrow} 0, \quad \frac{\tilde S_{1+}^2}{n \tilde S_{2+}} \overset{\P}{\rightarrow} 0, \quad \frac{\tilde S_{3+}}{n\tilde S_{2+}} \overset{\P}{\rightarrow} 0,
\end{align}
where $\overset{\P}{\rightarrow}$ denotes convergence in probability. For $\tilde w(|i-j|)$ obtained by centering the default weight \eqref{eqn:w(|i-j|)}, we have  
\begin{equation}
\frac{\tilde Z}{ \sqrt{ {\mathrm{var}}(\tilde Z) }} \stackrel{\mathcal{D}}{\rightarrow}  N(0,1) ~ \text{ when } n \rightarrow \infty
\end{equation}
under the permutation null distribution, where $\stackrel{\mathcal{D}}{\rightarrow}$ denotes convergence in distribution.
\label{thm: clt}
\end{theorem}

\begin{remark}
The result extends to other weights (Section~\ref{sec:weight schemes}) provided $\tilde w_{0+}\asymp \tilde w_{1+}\asymp 1$ and $\tilde w_{2+}\asymp \tilde w_{3+}\asymp n$. For example, we can choose $w(\cdot)$ to be either the geometric decay or the exponential decay weight function, as defined in Section \ref{sec:weight schemes}; both weight functions automatically satisfy the constraints.
\end{remark}

Theorem \ref{theory:normaltity} establishes asymptotic normality of $\tilde Z$ under the permutation null. It has the advantage that when conditional on the observed data, the only randomness comes from the random permutation, so $\mathrm{var}(\tilde Z)$ can be computed exactly. This typically yields a better finite-sample approximation to the sampling distribution of $\tilde Z$ than unconditional asymptotics, which require estimating the unknown variance $\mathrm{var}_0(\tilde Z)$ under $\mathsf{H}_0$ and thereby introduce additional estimation noise. For completeness, we also establish a Gaussian limit under the population null $\mathsf{H}_0$ (unconditional asymptotics) under mild moment conditions. Throughout, $\mathrm{E}_0$, $\mathrm{var}_0$, and $\mathrm{cov}_0$ denote the expectation, variance, and covariance under the population distribution respectively.

\begin{theorem}\label{thm:nullCLT}
Assume that $\E_0|\tilde S(X,X')|^{2+\delta}<\infty$ for some $\delta>0$ and i.i.d.\ samples $X,X'$. When $\tilde w(|i-j|)$ takes the centered form adapted from \eqref{eqn:w(|i-j|)}, it follows that under $\mathsf{H}_0$,
\begin{equation*}
\frac{\tilde Z}{ \sqrt{ {\mathrm{var}_0}(\tilde Z) }} \stackrel{\mathcal{D}}{\rightarrow}  N(0,1) ~ \text{ when } n \rightarrow \infty.
\end{equation*}
\end{theorem}

\subsection{Consistency analysis}
\label{sec:consistency}

We now proceed to investigate the power properties of WISE. To simplify the illustration, in what follows, let $\pi_0$ denote the identity permutation with $\pi_0(i) = i, \, i \in [n]$. Throughout, $C$ denotes a positive constant whose value may vary from line to line. 
For illustrative purposes, we focus on the case where the default weight function \eqref{eqn:w(|i-j|)} is used in Theorems~\ref{theorem: consistency under LDHSS} and~\ref{theorem: consistency under HDLSS}.

In the usual limiting regime where $n \to \infty$, the consistency of WISE is characterized by Theorem \ref{theorem: consistency under LDHSS}, which provides the theoretical foundation for the effectiveness of WISE in large-sample applications.
\begin{theorem}[Consistency under large-sample asymptotics]
\label{theorem: consistency under LDHSS}
    Let $\{X_t\}_{t=1}^n$ take value in a general probability space $(\Omega,\mathcal{F},\P)$.
    Assume that under $\mathsf{H}_a$, the following hold:
    \begin{itemize}
    
        \item[(C1)] $\{ X_t \}_{t=1}^{\infty}$ is a $\beta$-mixing strictly stationary process, with the mixing coefficient satisfying that for some $r>0$, $\beta(n) = O(n^{-r});$

        \item [(C2)] For some $\theta >0$ and some constant $C$ not dependent on $n$, $\sup_{1 \le i < j < \infty} 
        \mathrm{E}_0 [S_{ij}^{2+2\theta} ] \leq C;$
       
        \item[(C3)] $|\tr(\widetilde\W^{\top}\S)|/ \sqrt{n} \xrightarrow[]{\P} \infty$ when $n \to \infty$.

    \end{itemize}
    Then we have
    $$
    \mathrm{pr}  \Big( \Big| \frac{Z-\mathrm{E}(Z)}{\sqrt{\mathrm{var}(Z)}}  \Big| > z_{1- \frac{\alpha}{2}}|\mathsf{H}_a  \Big) \rightarrow 1 \quad \text{when~} n \rightarrow \infty,
    $$
    where $z_{1-\alpha/2}$ is the $(1-\alpha/2)$-quantile of the standard normal distribution.
\end{theorem}

\begin{remark}
    The $\beta$-mixing condition in (C1) is a standard and relatively mild assumption, widely used in the time series literature \citep{chang2017testing,jiang2024testing}. A broad class of commonly studied processes satisfies this property with the required decay rate, including causal autoregressive moving average (ARMA) processes with continuous innovation distributions, stationary GARCH processes with continuous innovation distributions and finite second moments, and stationary Markov chains. Further  examples can be found in \cite{fan2003nonlinear}. 

  Condition (C2) imposes a mild moment requirement, namely $\mathrm{E}_0 [S_{ij}^{2+2\theta}] \asymp 1$. When $\{X_t\}_{t=1}^n$ lie in a high-dimensional Euclidean space with dimension $p \to \infty$, this  can be satisfied by adopting a $p$-scaled similarity measure, for example $S(X_i, X_j) = - \|X_i - X_j\|_2/\sqrt{p}$. In particular, if $S(X_i,X_j)$ is the $p$-scaled Euclidean distance, a sufficient condition is  $\mathrm{E}_0 ( \|X_i\|_2^2 /p )^{1+\theta} \leq C$. 
\end{remark}

\begin{remark}
     The term $\tr(\widetilde{\W}^\top \S)$ in 
     (C3) is the Frobenius inner product  between $\S$ and $\widetilde{\W}$, which quantifies their structural alignment. Recalling the definition of  $w_{ij} $ in \eqref{eqn:w(|i-j|)}, $\widetilde{\W}$ encodes the expected similarity pattern under proximity-type dependence. Thus, $\tr(\widetilde{\W}^\top \S)$ directly measures how well the observed dependence structure matches the hypothesized proximity pattern: larger values indicating stronger agreement, while $\mathrm{E}_0(\tr(\widetilde \W^{\top} \S)) = 0$ when the observations are i.i.d.. For intuition, suppose that for some $\alpha>0$, $S_{ij} \asymp  \big[1 - 1/(1+|i-j|^{\alpha})\big]$ holds with probability $1-o(1)$. Then $|\tr(\widetilde\W^{\top}\S))|/n \xrightarrow[]{\P} c$ as $n \to \infty$, for some constant $c>0$. Additional empirical evidence supporting the plausibility of (C3) is provided in Supplementary~E.5.
\end{remark}

Although Theorem \ref{theorem: consistency under LDHSS} derives the consistency of WISE under the standard asymptotic regime, many modern data types, such as microarray data in genomics, brain imaging data in medical image analysis, and spectral measurements in chemometrics, typically arise in HDLSS settings. To address this, Theorem \ref{theorem: consistency under HDLSS} establishes the consistency of WISE in the HDLSS regime for proximity-type alternatives when the time series lie in Euclidean space and a specified similarity measure is employed.

\begin{theorem}[Consistency under HDLSS asymptotics]
\label{theorem: consistency under HDLSS}
Let $\{X_t\}_{t=1}^n \subset \R^p$ and define the similarity matrix as $\S^* =[S_{ij}^* ]_{i,j=1}^n =[- \sum_{k=1}^p(X_{ik}-X_{jk})^2/p ]_{i,j=1}^n$. 
Further assume that under $\mathsf{H}_a$, the following conditions hold:
\begin{itemize}
    \item[(C1')] $\{ X_t\}_{t=1}^{\infty}$ is weakly stationary process, with $\mathrm{E}_0(X_{tk}^4) \leq C$ for any $k \in [p]$ and some constant $C$ not dependent on $p$ and $n$;
    
    \item[(C2')] $\lim_{p \rightarrow \infty}\mathrm{E}_0 (S^*_{ij}) = \mu_{ij}$ exists, with $\mu_{ij}>\mu_{ik}$ for all $i \leq j <k$;
    
    \item[(C3')] Let
    $$
    \Delta_{ij} := \max_{u,v,r,s \in \{i,j \}} \sum_{1 \leq  k < l \leq  p} \Big|\mathrm{cov}_0(X_{u k} X_{v k}, \ X_{r l} X_{s l}) \Big |,
    $$
    and assume that $\max_{i,j \in [n]} \Delta_{ij} = o(p^2)$.
\end{itemize}
Then, for sufficiently large $n$,
$$
\mathrm{pr} \Big( Z > c_{1- \frac{\alpha}{2}}|\mathsf{H}_a \Big) \rightarrow 1, 
\quad \text{when } p \rightarrow \infty, 
$$
where $c_{1-\alpha/2}$ is the $(1-\alpha/2)$-quantile of the permutation null distribution of $Z$.
\end{theorem}

\begin{remark}
\label{remark:VAR1}
    Condition (C2') is imposed to guarantee that the serial dependence diminishes as the time gap increases, a property satisfied by a variety of proximity-type alternatives. For example, we consider a VAR(1) model specified as
\begin{equation*}
    \mathsf{H}_{a}^*: X_t = \A X_{t-1}+\varepsilon_t ,
\end{equation*}
where $\A \in \R^{p \times p}$ and $\{ \varepsilon_t \}_{t=1}^n$ are i.i.d. random vectors. It can be shown that (C2') will be satisfied if the following conditions hold:
\begin{itemize}
  \item[(C4')] For a positive definite matrix $\A$, $0<C_1 \leq \lambda_i(\A) \leq C_2 < 1$ for all $i \in [p]$, where constants $C_1$ and $C_2$ do not depend on $n$ and $p$, and $\{\lambda_i(\A)\}_{i=1}^p$ denotes the eigenvalues of $\A$. 
  
  \item[(C5')] $\tr((\mathbf{I}_p-\A^2)^{-1}\mathbf{\Sigma})/p \to C$, as $p \to \infty$,
  with $\mathbf{\Sigma} = \mathrm{var}_0(\varepsilon_t).$
\end{itemize}
Discussions about the verification of Remark 5 are in Supplementary E.7.

\end{remark}

\begin{remark}
    In the literature on HDLSS analysis, conditions analogous to (C3') are commonly adopted to ensure that the cross-sectional dependence of $X_t$ is relatively weak so that the available information can diverge as $p \to \infty$. For example, to derive the geometric representation of high-dimension-low-sample size (i.i.d.) data, \cite{hall2005geometric} supposed that all coordinates of $X_t$ form a cross-sectional series which is $\rho$-mixing for functions dominated by quadratics. A weaker version of this mixing condition was proposed by \cite{ahn2007high}, which only required $\rho$-mixing after a suitable permutation of the coordinates. Other mixing assumptions, such as $\alpha$-mixing, have also been employed for the same purpose; see, for instance, \cite{Li2018AsymptoticNO}.
\end{remark}

\section{Performance Analysis}\label{sec:Performance}

In this section, we conduct numerical experiments to evaluate the performance of our method in high-dimensional settings. By default, we define the similarity measure as $S(X_i, X_j) =  -\sum_{k=1}^p |X_{ik} - X_{jk}|$, and employ the weight function specified in  \eqref{eqn:w(|i-j|)}. The corresponding critical value is directly calculated according to Theorem \ref{theory:normaltity}.

To further assess the robustness of our approach, we also examine the performance of the test when alternative similarity measures and weight functions are used; detailed results are provided in Supplementary C.1 and C.2.

For comparison, we include four recently proposed methods: max-type test (MT, \cite{chang2017testing}); multivariate serial independence test based on the auto-distance correlation function (mADCF, \cite{fokianos2018testing}); rank-type test (RT, \cite{Tsay2020TestingSC}); CvM-type test (CvM, \cite{jiang2024testing}), a Cram\'er-von Mises-type statistic based on auto-distance covariance for random objects.
Additional details of these methods are summarized in Supplementary B.2. In the experiments, we implement MT at lags $l \in \{4,6,8,10\}$ (MT4, MT6, MT8, MT10), RT at lags $l \in \{1,5,10\}$ (RT1, RT5, RT10), CvM with the default Euclidean distance, and mADCF with the default truncated kernel and bandwidth $\sigma = 3n^{0.2}$, following the recommendations of \cite{fokianos2018testing} and \cite{jiang2024testing}. Since all methods except RT lack analytical critical values, we use $400$ (wild) bootstrap replications to approximate their critical thresholds. All simulation results are based on at least $350$ replications and are reported at the nominal significance level $\alpha = 0.05$.

We generate multivariate time series $\{X_t\}_{t=1}^{n} = \{(x_{t1},\dots,x_{tp})\}_{t=1}^n \subset \mathbb{R}^p$ under a variety of scenarios summarized in Table~\ref{Tab:simsetting}, with dimension $p \in \{200,400,800\}$ and sample size $n \in \{50,100,150\}$. Under the null hypothesis, the data are i.i.d. from four representative distributions (Settings 1.1-1.4), chosen to span a wide spectrum from cross-sectionally independent to dependent, from light-tailed to heavy-tailed, and from symmetric to skewed.

Under the alternatives, we design scenarios (Settings 2-5) that capture distinct dependence structures, ranging from linear correlations to uncorrelated nonlinear dependencies. These include proximity-type structures, such as VAR(1) and GARCH(1,1), as well as mixed-type structures, such as seasonal vector auto-regression (SVAR). For each setting, the signal-to-noise ratio is calibrated so that the best-performing method achieves moderate power, thereby avoiding trivial outcomes and ensuring a meaningful comparison.
\begin{table}[!htbp]
    \centering
    \renewcommand{\arraystretch}{1}
    \caption{Summary of simulation settings.}
    \vspace{5pt}
    \label{Tab:simsetting}
    \resizebox{\textwidth}{!}{%
    \begin{tabular}{ccc}
    \toprule
    Structure & Setting & Data Generation \\
    \midrule
    \multirow{4}{*}{i.i.d.} 
         & 1.1  & Multivariate Normal distribution\\
         & 1.2  & Multivariate Normal distribution with cross-sectional dependence  \\
         & 1.3  & Multivariate $t_1$ distribution  \\
         & 1.4  & Multivariate log-normal distribution \\
    \midrule
    \multirow{2}{*}{correlated} 
         & 2.1-2.3 & Vector auto-regression (VAR) \\
         & 3.1-3.2 & Seasonal vector auto-regression (SVAR) \\
         \midrule
         \multirow{2}{*}{\shortstack{dependent \\ but uncorrelated}} & 4  & Generalized auto-regressive conditional heteroskedasticity model (GARCH) \\
         & 5  & Nonlinear moving average (NMA) \\
    \bottomrule
    \end{tabular}%
    }
\end{table}

Before presenting the settings, we introduce notations that will be frequently used in later contexts. Let $\A$, $\B\in \R^{p\times p}$ be fixed matrices, $\mathbf{I}_p$ the $p$-dimensional identity matrix, and $\epsilon_t = (\epsilon_{t,1}, \dots, \epsilon_{t,p}) \in \R^p$, $t \in [n]$. Under null, the data is generated as follows:

\begin{enumerate}[label={}]
    \item (Setting 1): $X_t = \epsilon_t$, with $\epsilon_{t}$ i.i.d. generated.
    
    \begin{enumerate}[label={}]
    \item (Setting 1.1): $\epsilon_{t}\sim N(0,\mathbf{I}_p)$.
        
    \item (Setting 1.2): $\epsilon_{t}\sim N(0,\mathbf{\Sigma})$, where $\Sigma_{ij}=0.6^{|i-j|}$.
        
    \item (Setting 1.3): $\epsilon_{t}\sim \textup{Multivariate } t_{1}(0,\mathbf{I}_p)$.
        
    \item (Setting 1.4): $\epsilon_{t}\sim \exp(N(0,\mathbf{I}_p))$. 
    \end{enumerate}
\end{enumerate}

\begin{table}[t]
    \caption{Empirical sizes under Settings 1.1-1.4 for a nominal significance level of 0.05, with $n=50$ and $p \in \{200, 400, 800\}$.}
    \label{Tab:size50}
    \vspace{5pt}
    \centering
    \renewcommand{\arraystretch}{1} 
    \resizebox{\textwidth}{!}{ 
    \begin{tabular}{c c c c c ccc cccc}
    \toprule
        \multirow{2}{*}{Settings} & 
        \multirow{2}{*}{$p$} &
        \multirow{2}{*}{WISE} & 
        \multirow{2}{*}{CvM} & 
        \multirow{2}{*}{mADCF} &
        \multicolumn{3}{c}{RT} &
        \multicolumn{4}{c}{MT} \\ 
        \cmidrule(lr){6-8} \cmidrule(lr){9-12}
        & & & & & $l=1$ & $l=5$ & $l=10$ & $l=4$ & $l=6$ & $l=8$ & $l=10$ \\ 
    \midrule
        \multirow{3}{*}{1.1} & 200 & 0.050 & 0.060 & 0.000 & 0.017 & 0.011 & 0.007 & 0.000 & 0.000 & 0.000 & 0.000 \\
         & 400 & 0.052 & 0.054 & 0.000 & 0.021 & 0.012 & 0.010 & 0.000 & 0.000 & 0.000 & 0.000 \\
         & 800 & 0.062 & 0.062 & 0.000 & 0.020 & 0.011 & 0.005 & 0.000 & 0.000 & 0.000 & 0.000 \\
    \midrule
        \multirow{3}{*}{1.2} & 200 & 0.046 & 0.064 & 0.000 & 0.020 & 0.011 & 0.011 & 0.000 & 0.000 & 0.000 & 0.000 \\
         & 400 & 0.048 & 0.061 & 0.000 & 0.021 & 0.013 & 0.012 & 0.000 & 0.000 & 0.000 & 0.000 \\
         & 800 & 0.063 & 0.066 & 0.000 & 0.026 & 0.008 & 0.011 & 0.000 & 0.000 & 0.000 & 0.000 \\
    \midrule
        \multirow{3}{*}{1.3} & 200 & 0.050 & 0.064 & 0.032 & 0.015 & 0.010 & 0.008 & 0.012 & 0.012 & 0.014 & 0.014 \\
         & 400 & 0.059 & 0.062 & 0.029 & 0.011 & 0.009 & 0.013 & 0.008 & 0.008 & 0.008 & 0.009 \\
         & 800 & 0.051 & 0.056 & 0.031 & 0.014 & 0.009 & 0.009 & 0.016 & 0.015 & 0.010 & 0.013 \\
    \midrule
        \multirow{3}{*}{1.4} & 200 & 0.050 & 0.059 & 0.000 & 0.016 & 0.012 & 0.008 & 0.013 & 0.012 & 0.012 & 0.012 \\
         & 400 & 0.051 & 0.055 & 0.000 & 0.013 & 0.010 & 0.013 & 0.022 & 0.022 & 0.025 & 0.025 \\
         & 800 & 0.060 & 0.076 & 0.000 & 0.013 & 0.008 & 0.008 & 0.032 & 0.031 & 0.032 & 0.033 \\
    \bottomrule
    \end{tabular}}
\end{table}

The empirical sizes for $n = 50$ and $n = 100$ are reported in Tables~\ref{Tab:size50} and \ref{Tab:size100}, with the corresponding results for $n=150$ provided in Supplementary B.1. Across all settings, WISE exhibits accurate Type-I error control, with empirical sizes consistently close to the nominal level 0.05. For example, even in the most challenging case with $p=800$ and heavy-tailed distributions (Setting 1.3), the empirical size of WISE remains around $0.05$-$0.06$, demonstrating the reliability of our analytical critical formula in small samples.

The CvM test also achieves reasonable size control in most scenarios, though a slight inflation is observed. For example, under Setting 1.3 with $n=100$, the empirical size reaches $0.078$. Supplementary B.1 shows that this inflation diminishes as the sample size increases, except in heavy-tailed settings where deviations persist. This behavior is consistent with the reliance of CvM-type test on moment conditions, and thus aligns with theoretical expectations. By contrast, the other methods (MT, mADCF, RT) tend to be conservative in high-dimensional regimes, with empirical sizes rarely exceeding $0.03$. This under-rejection likely arises from the estimation of long-run covariance matrices (in MT and mADCF) and the sensitivity of rank-based statistics (in RT) when $p \gg n$.

Together, these findings highlight the robustness of WISE across diverse null distributions, while also illustrating the different regimes where each competing method encounters challenges.

\begin{table}[t]
    \caption{Empirical sizes under Settings 1.1-1.4 for a nominal significance level of 0.05, with $n=100$ and $p \in \{200, 400, 800\}$.}
    \label{Tab:size100}
    \vspace{5pt}
    \centering
    \renewcommand{\arraystretch}{1} 
    \resizebox{\textwidth}{!}{
    \begin{tabular}{c c c c c ccc cccc}
    \toprule
        \multirow{2}{*}{Settings} & 
        \multirow{2}{*}{$p$} &
        \multirow{2}{*}{WISE} & 
        \multirow{2}{*}{CvM} & 
        \multirow{2}{*}{mADCF} &
        \multicolumn{3}{c}{RT} &
        \multicolumn{4}{c}{MT} \\ 
        \cmidrule(lr){6-8} \cmidrule(lr){9-12}
        & & & & & $l=1$ & $l=5$ & $l=10$ & $l=4$ & $l=6$ & $l=8$ & $l=10$ \\ 
    \midrule
        \multirow{3}{*}{1.1} & 200 & 0.052 & 0.062 & 0.000 & 0.023 & 0.018 & 0.015 & 0.000 & 0.000 & 0.000 & 0.000 \\
         & 400 & 0.055 & 0.064 & 0.000 & 0.018 & 0.022 & 0.022 & 0.000 & 0.000 & 0.000 & 0.000 \\
         & 800 & 0.051 & 0.061 & 0.000 & 0.020 & 0.020 & 0.019 & 0.000 & 0.000 & 0.000 & 0.000 \\
    \midrule
        \multirow{3}{*}{1.2} & 200 & 0.046 & 0.062 & 0.000 & 0.025 & 0.020 & 0.019 & 0.000 & 0.000 & 0.000 & 0.000 \\
         & 400 & 0.048 & 0.077 & 0.000 & 0.038 & 0.020 & 0.012 & 0.000 & 0.000 & 0.000 & 0.000 \\
         & 800 & 0.037 & 0.066 & 0.000 & 0.024 & 0.024 & 0.023 & 0.000 & 0.000 & 0.000 & 0.000 \\
    \midrule
        \multirow{3}{*}{1.3} & 200 & 0.051 & 0.070 & 0.021 & 0.026 & 0.017 & 0.016 & 0.007 & 0.008 & 0.008 & 0.008 \\
         & 400 & 0.057 & 0.078 & 0.021 & 0.022 & 0.017 & 0.019 & 0.003 & 0.004 & 0.006 & 0.007 \\
         & 800 & 0.058 & 0.069 & 0.014 & 0.025 & 0.024 & 0.008 & 0.001 & 0.001 & 0.006 & 0.001 \\
    \midrule
        \multirow{3}{*}{1.4} & 200 & 0.050 & 0.057 & 0.000 & 0.028 & 0.018 & 0.016 & 0.000 & 0.000 & 0.000 & 0.000 \\
         & 400 & 0.050 & 0.056 & 0.000 & 0.025 & 0.012 & 0.015 & 0.001 & 0.001 & 0.001 & 0.001 \\
         & 800 & 0.055 & 0.067 & 0.000 & 0.022 & 0.014 & 0.011 & 0.000 & 0.000 & 0.000 & 0.000 \\
    \bottomrule
    \end{tabular}}
\end{table}

The data generation schemes for the alternatives are  as follows:
\begin{enumerate}[label={}]
    \item (Setting 2): VAR(1). $X_t = \mathbf{A} X_{t-1} +\epsilon_t$, 
    with $\epsilon_{t}\overset{i.i.d.}\sim N(0,\mathbf{I}_{p})$. We consider different levels of sparsity for $\mathbf{A}$ in the following sub-settings: 
    \begin{enumerate}[label={}]
        \item (Setting 2.1): $\A = 0.015\mathbf{I}_p$.               
        \item (Setting 2.2): $A_{ij}$ are independently generated from $ \text{U}(-0.01,0.04) $ for $|i-j|\leq p/50$, and set to $0$ otherwise.
        \item (Setting 2.3): $A_{ij}$ are independently generated from $ \text{U}(-0.04,0.015) $ for $|i-j|\leq p/20$, and set to $0$ otherwise. 
    \end{enumerate}

    \item (Setting 3): SVAR(1,1). $X_t = \A X_{t-l}+\B X_{t-1}-\A \B X_{t-l-1}+\epsilon_t$, where $\epsilon_{t}\overset{i.i.d.}\sim N(0,\mathbf{I}_{p})$ and $l$ is the seasonal lag order. Here we define $\mathbf{A}$ and $\mathbf{B}$ as follows: $A_{ij} \sim \text{U}(-0.01, 0.03)$ and $B_{ij} \sim \text{U}(-0.01, 0.04)$ independently for $|i-j| \leq p/50$, and $A_{ij} = B_{ij} = 0$ otherwise. We consider two seasonal orders: $l=4$ (Setting 3.1) and $l=12$ (Setting 3.2). 
    
    \item (Setting 4): GARCH(1,1).  
    $X_{t} = h_{t}\circ \epsilon_{t}$, with
    $ h_t^{\circ 2}= b + \A X_{t-1}^{\circ 2} + \B h_{t-1}^{\circ 2}$,
    $\epsilon_{t}\overset{i.i.d.}{\sim}N(0,\mathbf{I}_p)$, and $b = (0.002,\dots,0.002)\in \mathbb{R}^p$. Both $\mathbf{A}$ and $\mathbf{B}$ are diagonal matrices, with $A_{ii}$ and $B_{ii}$ independently generated from $\text{U}(0,0.15)$ and $\text{U}(0,0.4)$, respectively.

    \item (Setting 5): NMA(2). $X_{t} = \epsilon_{t} \circ \epsilon_{t-1}\circ\epsilon_{t-2}$,  $\epsilon_{t}\overset{i.i.d.}{\sim}N(0,\mathbf{I}_p)$.
\end{enumerate}

\begin{figure}[t]
    \setlength{\abovecaptionskip}{1pt} 
    \centering
    \includegraphics[width=1\linewidth]{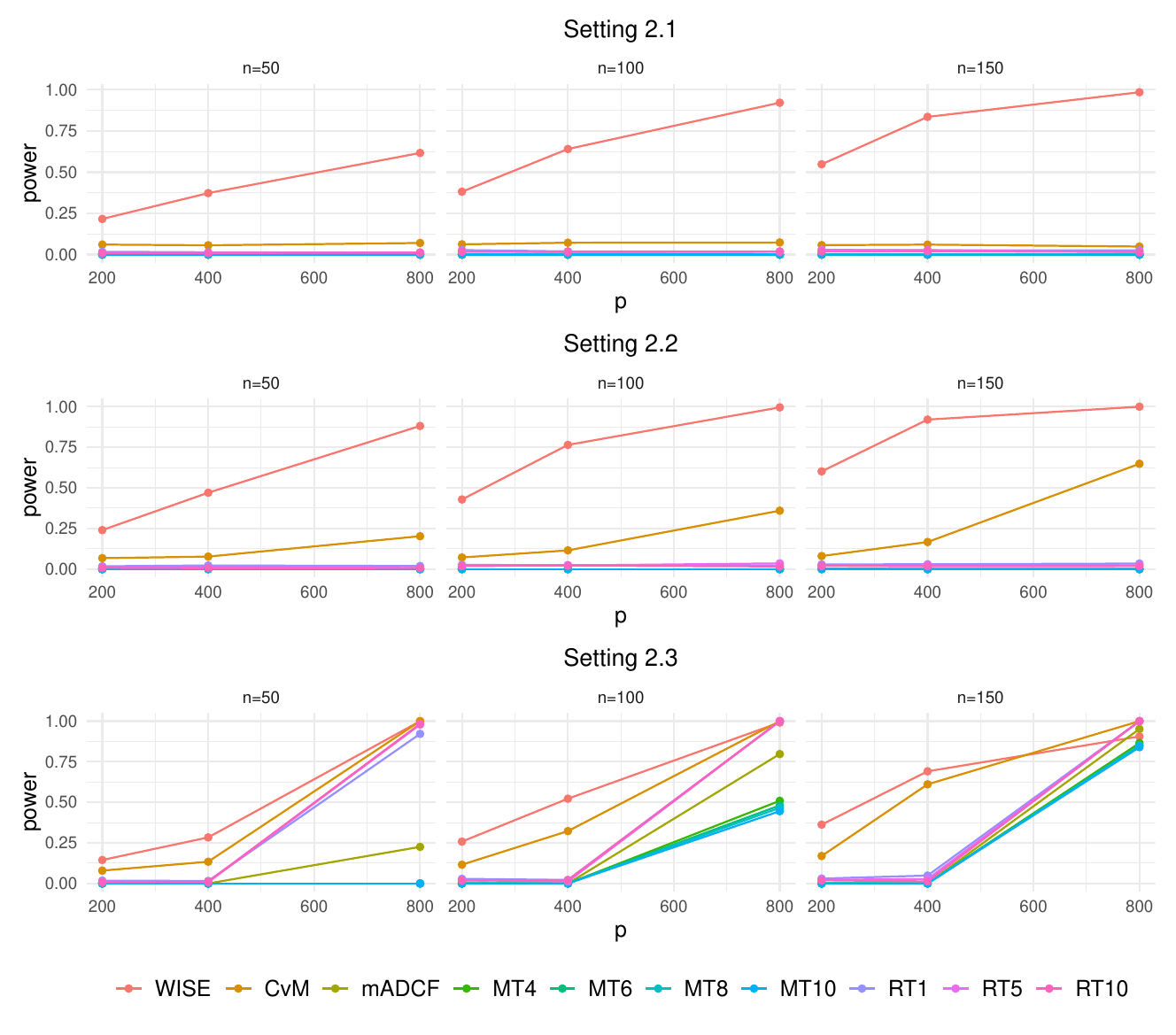}
    \caption{Empirical power for Settings 2.1 to 2.3, with sample size $n \in \{50,100,150\}$ and dimension $p \in \{200,400,800\}$.}
    \label{fig:setup2}
    \end{figure}

\begin{figure}[t]
    \setlength{\abovecaptionskip}{1pt}
    \centering
    \includegraphics[width=1\textwidth]{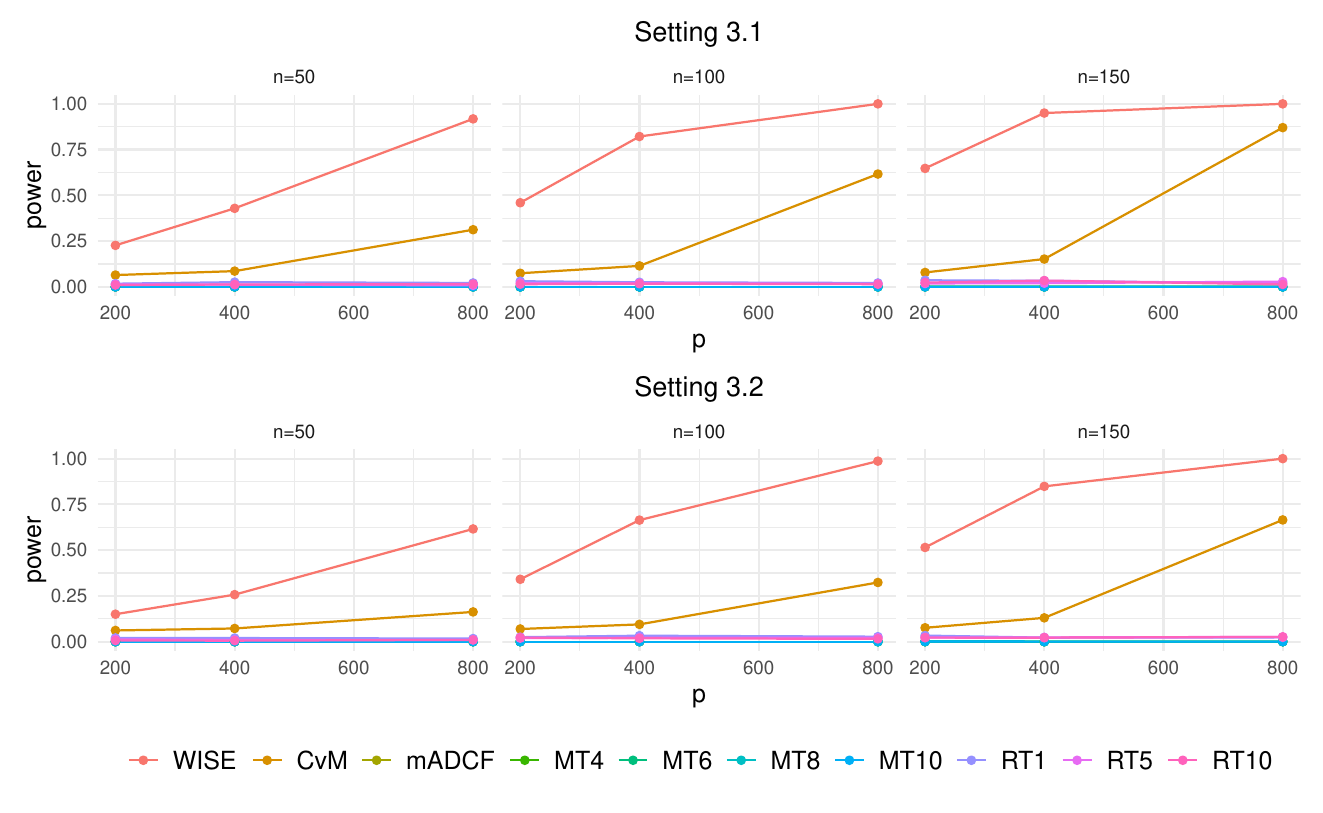}
    \caption{Empirical power for Settings 3.1 to 3.2, with sample size $n \in \{50,100,150\}$ and dimension $p \in \{200,400,800\}$.}
    \label{fig:setup3}
\end{figure}

\begin{figure}[t]
    \setlength{\abovecaptionskip}{1pt}
    \centering
    \includegraphics[width=1\textwidth]{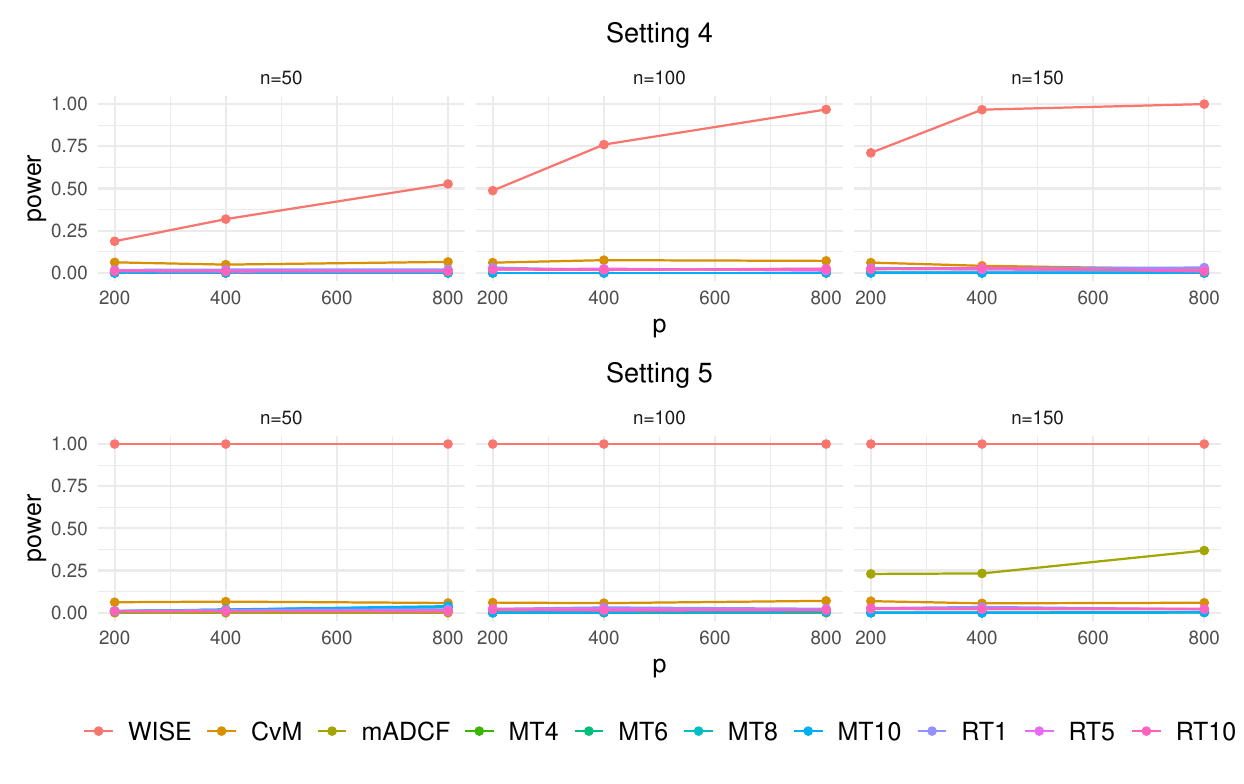}
    \caption{Empirical power for Settings 4 and 5, with sample size $n \in \{50,100,150\}$ and dimension $p \in \{200,400,800\}$.}
    \label{fig:setup45}
\end{figure}

The empirical power results for all methods under Settings 2-5 are displayed in Figures~\ref{fig:setup2}-\ref{fig:setup45}. Overall, WISE demonstrates consistently strong performance across these alternatives. The only exception is Setting 2.3 with $(n,p)=(150,800)$, where CvM, mADCF, and RT slightly surpass WISE; however, the difference in power is less than 0.1. Importantly, in the uncorrelated dependence structures of Settings 4 and 5, WISE maintains substantial power, whereas the rejection rates of the other methods remain low. This highlights the ability of the proposed test to capture non-linear and uncorrelated dependence in ultra-high-dimensional contexts.

Among the competing approaches, CvM shows relatively good performance under correlated alternatives such as Settings 2 and 3, often ranking second to WISE. In contrast, its power is limited in uncorrelated but dependent scenarios (Settings 4 and 5), where its rejection rates rarely exceed $0.1$. This phenomenon is in line with previous observations in \cite{ramdas2015decreasing} and theoretical insights in \cite{zhu2020distance}, which showed that the distance covariance tends to degenerate into the sum of squared componentwise sample cross-covariances when $p \gg n$.

The other methods (mADCF, MT, RT) achieve noticeable power only in selected settings. For example, MT and RT perform well in Setting 2.3, where the correlation strength is relatively high, but their performance is weaker in other scenarios. This behavior aligns with the methodological focus of these tests, which are more effective when strong correlation signals are present.

In summary, the simulation results demonstrate that WISE provides robust detection ability across a broad range of dependence structures, including challenging cases with uncorrelated and non-linear dependence. At the same time, the comparison highlights the complementary nature of existing approaches: CvM remains competitive in correlated regimes, while MT, mADCF, and RT retain advantages in scenarios that align with their design assumptions. In addition, we conduct further numerical experiments, reported in Supplementary C, to assess the performance of our test on bivariate time series. The results indicate that the test maintains accurate size control and competitive power in low-dimensional settings.

\section{Analysis of Tokyo Check-ins Data}
\label{sec:RealData}

Location-based social networks (LBSNs) provide a novel source of spatio-temporal data that captures the social behaviors of mobile users. In LBSNs, each activity is typically recorded as a check-in at a physical venue, such as a bus station, bank, or coffee shop—together with its GPS coordinates, visit time, and venue category. With the rapid growth of mobile internet and wearable devices, such data have been widely used to study the spatio-temporal patterns of human mobility \citep{Ding2018SpatialTemporalDM, Li2021ExploringTA, Wang2022ModelingSN}. Previous studies suggest that check-in behaviors often exhibit temporal regularity, in the sense that users tend to revisit similar locations within a given time frame \citep{Yang2015ModelingUA}. In the following, we conduct a statistical analysis to examine whether such temporal consistency can be formally detected.

\begin{figure}[!t]
    \centering
    \includegraphics[width=\textwidth]{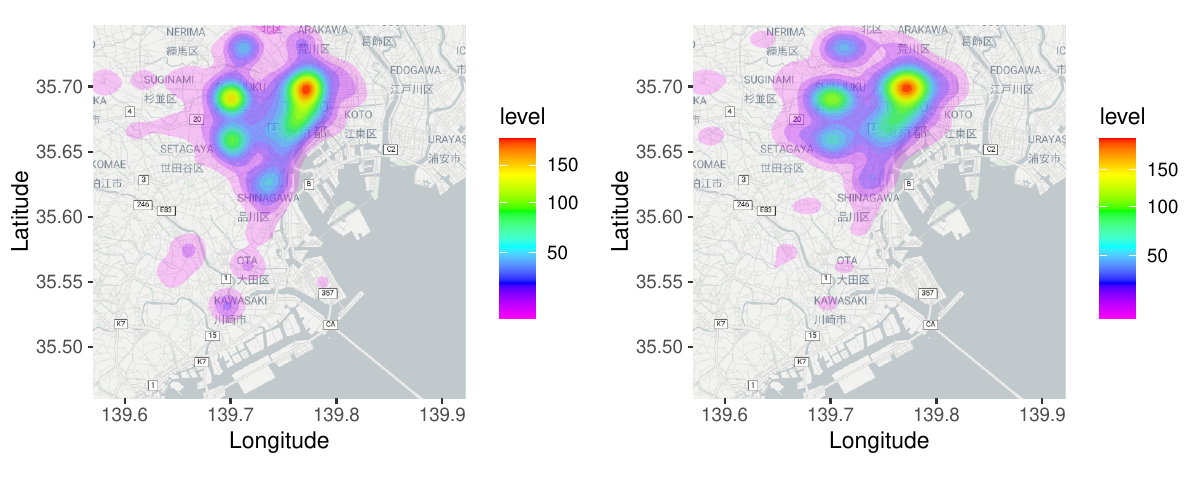}
    \vspace{-30pt}
    \caption{Density heatmaps of check-ins in Tokyo for April 4th, 2012 (left) and April 5th, 2012 (right).}
    \label{fig:heatmaps for check-ins}
\end{figure}

The Tokyo Check-ins Dataset\footnote{\href{https://www.kaggle.com/datasets/chetanism/foursquare-nyc-and-tokyo-checkin-dataset}{https://www.kaggle.com/datasets/chetanism/foursquare-nyc-and-tokyo-checkin-dataset}} contains over 570,000 check-in records from 1,939 mobile users in Tokyo, covering the period from April 2012 to February 2013. To illustrate the geographical distribution of check-ins, we visualize data recorded over six consecutive days, from April 4 to April 9, 2012. Figure~\ref{fig:heatmaps for check-ins} shows the check-ins for the first two days, while the remaining days are presented in Supplementary D. Notably, check-ins are concentrated in similar locations across different days, suggesting that users may exhibit stable preferences for specific regions within a given time frame. To enable statistical testing of such potential spatial preferences, we carry out the following preprocessing steps: 
\begin{enumerate}
    \item The geographical extent of Tokyo is restricted to latitudes between $35.5$ and $35.9$ and longitudes between $139$ and $140$, and partitioned into a $20\times20$ grid of equal-sized cells. 
    
    \item The number of check-ins in each cell is counted for every day, so that each day is represented by a $20 \times 20$ matrix of counts. 
\end{enumerate}

We consider our proposed test as well as four competing methods: CvM, mADCF, MT at lag 2 (MT2), and RT at lag 2 (RT2). For implementation, the similarity measure in our method is defined as the negative Frobenius norm, while CvM also adopts the Frobenius norm between matrices as its metric. Since MT, RT, and mADCF cannot be applied directly to matrix-type data, we reshape the original matrix series into high-dimensional vectors, eliminate zero-only columns, and set the remaining tuning parameters for these methods to the same values as in Section \ref{sec:Performance}.

\begin{table}[!htp]
\centering 
\caption{Decisions of five methods for the Tokyo check-ins dataset at nominal level of $0.05$ and $0.01$. Here $1$ denotes rejection, $0$ denotes non-rejection.} 
\vspace{5pt}
{
\renewcommand{\arraystretch}{0.9} 
\begin{tabular}{c cc cc cc cc cc} 
\toprule 
\multirow{2}{*}{Period} & \multicolumn{2}{c}{WISE} & \multicolumn{2}{c}{CvM} & \multicolumn{2}{c}{mADCF} & \multicolumn{2}{c}{MT2} & \multicolumn{2}{c}{RT2}\\ 
\cmidrule(lr){2-11} 
& $0.05$ & $0.01$ & $0.05$ & $ 0.01$ & $0.05$ & $ 0.01$ & $0.05$ & $ 0.01$ & $0.05$ & $ 0.01$\\ 
\midrule 
May-Jul & $1$ & $1$ & $1$ & $0$ & $0$ & $0$ & $0$ & $0$ & $0$& $0$\\
Aug-Oct & $1$ & $1$ & $1$ & $0$ & $0$ & $0$ & $0$ & $0$ & $0$ & $0$\\
\bottomrule 
\end{tabular}
}
\label{Tab: TKY}
\end{table}
We apply tests to two distinct time periods: May-July 2012 and August-October 2012. For illustrative purposes, we treat these as two separate tests rather than as a multiple testing problem. The results are summarized in Table~\ref{Tab: TKY}, which reports only the decisions of the five tests, since the CvM-type test does not provide the exact $p$-value. We observe that, both the proposed test and CvM reject the null hypothesis at the $0.05$ significance level, indicating a strong temporal consistency in users' check-in behaviors. This finding aligns with the conclusions of \citet{Yang2015ModelingUA} as well as our exploratory data analysis. Moreover, the $p$-values of WISE are 0.0001 and 0.0008 for the May-July and August-October periods respectively, while CvM fails to reject the null at the 0.01 significance level in either period, suggesting that our method may offer greater power. However, the other methods, including MT2, RT2, and mADCF, fail to detect the temporal dependence in users' geographical activities, which is consistent with our observation in Section \ref{sec:Performance} that the power of these methods remains relatively limited in high-dimensional scenarios.

\section{Conclusion}
\label{sec:conc}
We have proposed a flexible framework for testing serial independence that is powerful across a wide range of high-dimensional settings. Under mild conditions, the procedure admits an analytic critical value and extends naturally to non-Euclidean data whenever an appropriate similarity measure is available.

Meanwhile, several aspects warrant further investigation to enhance the overall effectiveness of the proposed approach. For instance, based on different weight functions $w(|i-j|)$, we can construct various statistics $Z$ as defined in \eqref{eqn:double indexed}, with each tailored to capture diverse dependence structures. These statistics can then be combined into a Mahalanobis-type test statistic of the form:
\begin{equation*}
    M = (Z_1-\mu_1,\dots,Z_m-\mu_m)\Sigma^{-1}_M(Z_1-\mu_1,\dots,Z_m-\mu_m) ^{\top},
\end{equation*}
\noindent where $ \Sigma_M = \mathrm{cov}(Z_1-\mu_1,\dots, Z_m - \mu_m)$ and $ \mu_i = \mathrm{E}(Z_i)$, $i = 1,\dots,m$, with $\mathrm{cov}$ denoting the covariance under the permutation null distribution. For applications where periodicity-type alternatives are of interest, one could define the weight functions as $w^{(k)}_{ij} = \cos(2\pi |i-j|/h_k) - 1$ for $k = 1,\cdots,m$, which may enable $M$ to simultaneously detect multiple periodicities. We leave the systematic study of such aggregated statistics, together with their theoretical properties and empirical performance, as an avenue for future research.


\bibliographystyle{chicago}
\bibliography{Bibliography-MM-MC}

\end{document}